# An allometry-based approach for understanding forest structure, predicting tree-size distribution and assessing the degree of disturbance

**Tommaso Anfodillo[1*], Marco Carrer[1], Filippo Simini[2,3], Ionel Popa[4], Jayanth R. Banavar[5] and Amos Maritan[6]**

[1] Dipartimento Territorio e Sistemi Agro-Forestali, University of Padova, 35020 Legnaro, Italy.

[2] Center for Complex Network Research and Department of Physics, Northeastern University, Boston, Massachusetts 02115, USA.

[3] Institute of Physics, Budapest University of Technology and Economics, Budapest, H-1111, Hungary

[4] Forest Research and Management Institute, Campulung Moldovenesc, Romania.

[5] Department of Physics, University of Maryland, College Park MD 20742, USA.

[6] Dipartimento di Fisica "G. Galilei", University of Padova, CNISM and INFN, 35131 Padova, Italy.

Corresponding author:

Tommaso Anfodillo; e-mail: tommaso.anfodillo@unipd.it

**Summary**

Tree-size distribution is one of the most investigated subjects in plant population biology. The forestry literature reports that tree-size distribution trajectories vary across different stands and/or species, while the metabolic scaling theory suggests that the tree number scales universally as -2 power of diameter.

Here, we propose a simple functional scaling model in which these two opposing results are reconciled. Basic principles related to crown shape, energy optimization and the finite size scaling approach were used to define a set of relationships based on a single parameter, which allows us to predict the slope of the tree-size distributions in a steady state condition.

We tested the model predictions on four temperate mountain forests. Plots (4 ha each, fully mapped) were selected with different degrees of human disturbance (semi-natural stands *vs.* formerly managed).

Results showed that the size distribution range successfully fitted by the model is related to the degree of forest disturbance: in semi-natural forests the range is wide, while in formerly managed forests, the agreement with the model is confined to a very restricted range.

We argue that simple allometric relationships, at individual level, shape the structure of the whole forest community.

**Introduction**

Plants show a notable regularity of structures and functions with change in size, which can be successfully described through allometric relationships [1]. These regularities exist because, at an individual level, plants must seek fitness, and optimization principles shape plants in a self-similar manner [2]. However, the structure of single individuals also appears to be relevant for determining the properties at higher levels of organization (e.g. populations and communities) that emerge from the interactions of adaptive individuals with each other and with their surroundings [3]. One of the most studied properties of tree communities is the shape of the tree-size distribution (i.e. the self-thinning line). Because ecosystem structure is tightly linked to its functionality, the tree-size distribution plays an essential role as indicator of community status and for inferring the role of forests in the global carbon budget [4,5]. A detailed analysis of the community structure appears to be even more important for implementing a close-to-nature silviculture, i.e. a forest management scheme aimed at mimicking all natural processes governing growth and mortality of trees, including competition and disturbances [6].

New insights into tree-size distributions were proposed by deriving the predictions of the community structure from the relationships based on metabolism and allometry at individual level, (i.e. "the forest is the tree") [2,7]. Formerly, this information had been sought either through empirical approaches or complex process-based models [e.g. 8]. Both these methods were often unsuitable for management purposes due to data needs and parameterization requirements [9]. West *et al.* [2] and Enquist *et al.* [7] have proposed a fascinating generalization that opens new perspectives in dealing with the complexity of forest ecosystems and suggests that relatively simple deterministic rules might be invoked for explaining their structure when the community is fully saturated or, equivalently, it uses all available resources. Strikingly, their scaling approach simply predicts that the self-thinning line should be a general property of forests across the globe (both evenly and unevenly aged) and ought to converge towards a slope of -2 (i.e. $N_{max} \propto d^{-2}$ where $N_{max}$ is the number of individuals in a given diameter class and $d$ is the tree diameter class). Yet, many foresters and ecologists have criticized this unifying approach mostly because the forestry literature is full of evidence that the slopes of the self-thinning lines vary significantly with different species, forest stands and, clearly, in relation to disturbance events [10-13].

The aim of this paper is to answer the following three questions: i) is it possible, by using simple allometric relationships at individual tree level, to predict the structure (i.e. tree-size distribution) of the whole community in a virtual steady state condition? ii) does the slope of the tree-size distribution curve differ among forest stands? iii) can the degree of disturbance be assessed comparing the actual tree distribution with the predicted structure of the fully saturated community? Here we present a general approach and an application to four temperate mountain forests with different disturbance regimes, where all the data needed (in particular tree height, crown length and crown radius) were systematically collected on more than 12,000 trees.

## Methods

### The "H-model" and the finite-scaling approach

Our model ("$H$-model" in the following) simply assumes [14] that the projected crown width (i.e. crown radius, $r_{cro}$ ) versus tree height ($h$) has a power law form $r_{cro} \propto h^H$. When $H=1$, the crown shape does not change with ontogenesis; if $H<1$ the crown becomes proportionally narrower in adult trees. Assuming a constant foliage density with ontogenesis, it follows that the crown leaf mass (or area) scales as the crown volume ($V_{cro}$), thus $V_{cro} \propto h^{1+2H}$. Clearly, the amount of foliage can be very different among species (e.g. shade-tolerant species have a larger leaf mass than shade-intolerant species for similar $V_{cro}$) but the *relative* variation with ontogenesis might scale similarly. We further assume that the metabolic rate ($B$) scales isometrically with leaf area (i.e. $B \propto V_{cro}^1$) as already demonstrated [15], leading to $B \propto h^{1+2H}$. The isometry between metabolic activity and leaf area is preserved with tree size given that growing trees are able to compensate for the increase in the distance from roots to leaves by tapering xylem conduits [16-18]. Moreover, the tree distribution network is not required to be fractal-like, thus overcoming one of the most common criticisms of the West *et al.* [2] model as discussed by Duursma *et al.* [19].

At the community level we hypothesize that tree-size distribution in natural ecosystems is *not* a pure power law but rather it follows *finite-size scaling* [20]. Due to resource limitations, there is a limit on the largest observable tree (i.e. the maximum tree height in a given site). If the range of possible sizes is wide enough, there is a power law regime followed by a sharp decrease, on approaching the upper limit, in the number of individuals (see Appendix 1). The uppermost value of the power law behaviour has been called the *characteristic height* ($h_c$). Thus, for a given value of the characteristic height $h_c$, we assume that the probability of finding a tree taller than *h*, i.e. the cumulative distribution of heights, follows the finite size scaling relationship:

$$P^>\big(h|h_c\big) = h^{-\alpha} f_h\big(h/h_c\big)$$

(eqn 1)

where $\alpha$ represents the exponent of the pure power law (i.e. the exponent of the self-thinning line) and the second term $f_h\big(h/h_c\big)$ tends to a constant when $h << h_c$, thus leading to pure power law behaviour, whereas it declines rapidly to zero when *h* approaches $h_c$, with a consequent deterioration of the quality of a power law fit. Many forms of the function $f_h\big(h/h_c\big)$ may be chosen. For the purposes of the analysis in this paper, we empirically chose a form that clearly underscores both the lower and upper cut-offs and is entirely consistent with the form of Eq. (1). Note that the standard probability distribution function (PDF) is simply obtained by differentiation, i.e. $p(h|h_c)$=-$dP^>(h|h_c)/dh$.

Finally we assume that, as in West *et al.* [2], the community is able to maximize the use of all available resources and, in this optimal condition, the total metabolism of the whole forest is *proportional* to the volume filled by the forest, estimated as the area of the forest (A) times $h_c$; thus with the relationships described in Simini *et al.* [14] it follows that $P^>\big(h|h_c\big) \propto 1/h^{2H}$ for tree heights $h<h_c$.

Therefore, our model predicts that the fraction of trees corresponding to tree size ($h$) should scale as $-dP^>\left(h|h_c\right)/dh \propto 1/h^{1+2H}$ which represents exactly the self-thinning trajectory and scales similar to the maximum number, $N_{max,h}$, of plants of height $h$, that can grow on a given area. Notably, the exponent is the *same*, but with negative sign, as the scaling of crown volume with $h$. This indicates that the self-thinning line is simply driven by the scaling of the crown volume of the single trees.

We selected 4 different sites: 1) Romania – Slatioara (altitude 1100 m, 47.27 N, 25.80 E); 2) Italy – Cansiglio (altitude 1000 m, 46.06 N, 12.25 E); 3) Italy – Cortina d'Ampezzo (altitude 2100 m, 46.29 N, 12.06 E); 4) Italy – Obereggen (altitude 1900 m, 46.23 N, 11.32 E) (Table 1). These sites were chosen in order to compare similar forest types with different anthropogenic disturbances. The forest located in Romania is a mixed Silver fir (*Abies alba* Mill.), European beech (*Fagus sylvatica* L.) and Norway spruce (*Picea abies* (L.) Karst.) stand and is classified as a virgin forest. A similar forest type grows on site 2) Italy – Cansiglio, but it was managed until 1980 and the effects of human activity on the forest structure are still evident. The other two sites are mixed high-altitude conifer forests composed of European larch (*Larix decidua* Mill.), Cembran pine (*Pinus cembra* L.) and Norway spruce 3) Italy – Cortina d'Ampezzo and 4) Italy – Obereggen. The former is one of the most undisturbed forests on the Dolomites (no logging recorded in the last 150 years), while site 4) was managed until 1990 and still maintains a structure influenced by human activities.

In all forests a permanent plot of 4 ha was established and a common measurement protocol implemented. All trees taller than 130 cm were identified, labelled and the following features recorded: topographic position, species, diameter at breast height (D), total height ($h$), height of the lowest living branches, and four radii of the vertical crown projection in the two directions marked by the plot axes.

We used data for tree height ($h$), length of the crown ($l_{cro}$), averaged radius of the crown ($r_{cro}$) and diameter distribution. $V_{cro}$ is estimated to scale as $r_{cro}^2 * l_{cro}$. Next, we calculated the exponent ($H$) from the scaling relationship $V_{cro} \propto h^{1+2H}$. To quantitatively assess the stand disturbance level in each site, we fitted the experimental tree-size distributions by introducing a modified version of equation 1 (i.e. equation 2) that in addition to the upper cut-off at large heights (i.e. the characteristic height $h_c$) has a lower cut-off, $h_{inf}$, that allows one to take into account possible deviations from the power law regime at low heights.

$$P^{>}(h \,|\, h_c, h_{inf}) = \frac{1}{\left[1 + (h - h_0)/h_{inf}\right]^{2H}} e^{-\left(\frac{h}{h_c}\right)^{5.5} + \left(\frac{h_0}{h_c}\right)^{5.5}} . \qquad \text{(eqn 2)}$$

Using equation 2 it is possible to obtain a more precise estimate of the height range in which the power law regime holds. We chose a stretched exponential with stretching exponent 5.5 as the scaling function, f, because it was the simplest functional form that provided good agreement with the data in the rapid decay region, $h \gg h_c$. However, other forms for the scaling function are allowed, provided that they have no stand-specific free-parameter other than hc. The parameters have been fitted using the least-squares method. $h_0$ is the minimum tree height considered in the collected data and is introduced simply because $P^{>}$ has to become $1$ at $h=h_0$ and $h_c$ is the characteristic height appearing in equation 1. Since $P^{>}$ *is* dimensionless, a further reference height, $h_{inf}$, must necessarily be introduced in the pre-factor of equation 2. In the range where both $h_{inf}$ and $h_0$ are much smaller than $h$ equation 2 reduces to the cumulative distribution defined in equation 1 in which the scaling function, f, is a stretched exponential $f(h \,|\, h_c) = e^{-(h/h_c)^{5.5}}$ with stretching exponent set to 5.5 for all sites. Equation 2 has two important parameters: $h_c$ and $h_{inf}$, where $h_{inf}$ is the lowest tree height above which there is a power law distribution and $h_c$ is, as mentioned above, the upper cut-off of the power law regime; $h_0$ is the height of the smallest trees measured (1.3 m) and $H$ is the value of the scaling exponent estimated, in each site, by using equation 2. The normalized difference $I_r=(h_c-h_{inf})/h_{max}$, where $h_{max}$ is the dominant tree height in each site (averaging the tallest 40 trees in the plots), i.e. the normalized range over which power law behaviour is observed, defines the "*recovery index*", which quantitatively expresses the degree of ecosystem reorganization since disturbance: the higher the $I_r$, the higher the community old-growthness. Indeed, a large $I_r$ indicates that $P^>(h|h_c)$ has a wide range with power law behaviour *consistent* with the prediction of $H$ from equation 2, indicating full resource use.

**RESULTS**

The four forests appear to be rather different in terms of structural properties (table 1). In the virtually undisturbed stands (sites 1 and 3), the number of individuals is twice that of the similar but formerly managed stands. Moreover, the mean diameter is smaller in the undisturbed forests indicating that they are populated by a significantly larger number of small trees. Nevertheless, the largest diameters were recorded in the undisturbed sites: 143 cm in site 1 and 95 cm in site 3.

The scaling of crown volume with tree height was surprisingly similar in all forests and the exponents of the relationships varied only from 2.2 to 2.3 (figure 1). The values of the $H$ exponent, derived by using the relationship $(V_{cro}) \propto h^{1+2H}$ did not differ within similar forest types: site 1 (95% CI 0.59-0.63); site 2 (95% CI 0.62-0.68); site 3 (95% CI 0.61-0.65); site 4 (95% CI 0.63-0.68) but the intercepts (i.e. the $V_{cro}$ of a tree 130 cm tall) significantly changed among sites. The amount of foliage (in relative units) is one order of magnitude larger in mountain forests (sites 1-2) compared to high-altitude forests (sites 3-4). Within the same site, however, the scaling of the crown volume with tree height appeared to be very similar between species in spite of their different functional types (e.g. deciduous *versus* evergreen). The scaling exponents of Silver fir (2.26; 95% CI 2.20-2.30) and European beech (2.38; 95% CI 2.30-2.46) in site 1 and of European larch (2.21; 95% CI 2.16-2.24) and Cembran pine in site 3 (2.29; 95% CI 2.24-2.36) (figure 2) were not significantly different.

The ability of equation 2 in fitting the CDFs (i.e. the self-thinning curve) is shown in figure 3 and the estimated parameters are summarized in Table 2. In sites 1 and 3, by summing the leaf area of the smallest trees until $h_c$ it appeared that about 75-80% of the total community leaf area and about 95% of the total number of trees are accounted for at that threshold. This means that trees with height above the 95[th] percentile had about a quarter of the total leaf area of the whole community. The recovery index ($I_r$) is higher in the more natural forests (sites 1 and 3) and the range of the CDFs consistent with the relationship $P_h(h|h_c) \propto h^{-(1+2H)}$ is much wider, thus indicating that the self-thinning trajectories (i.e. the community structures) follow the predictions derived from the allometry of individual trees.

**Discussion**

*Methodological issues*

We would suggest a different approach to analysing the tree-size distribution curves by promoting the awareness that the concept of *finite size scaling* represents a key issue. Indeed, all the tree-size distributions published until now clearly depart from the power law in the regime of larger sizes [e.g. 2,5,13,21,22]. We propose an alternative explanation for the fact that the number of trees with large diameter/height is lower than that predicted by a power law scaling – interestingly, this explanation is related neither to a higher mortality in big trees or reduction in nutrient availability as previously suggested [22,23] nor because deaths of trees in older stands create large gaps that are slow to refill [24]. Instead, it could be a simple consequence of resource limitations, which lead to a limited size (e.g. a maximum tree height in a given site), as we demonstrated with a numerical example (see Appendix 1). It is entirely possible that all these factors to varying degrees play a role in a real forest. In any case, in forest communities, pure power law behaviour can only hold over a limited range of sizes. Without taking into account the finite size scaling in the fitting procedure, there is necessarily some arbitrariness in the determination of the power law exponent, which can yield unrealistically high slopes < -3.0 [5].

*Tree height and the Energy Equivalence Rule* (EER)

The importance of using tree height rather than tree diameter as independent variable emerges for its higher predictive skill of the properties of forest communities (e.g. productivity) as pointed out by Kempes *et al*. [25].

In this regard, one of the most relevant properties of a forest community is the near constancy of the cumulative leaf area during the self-thinning process [26].

This "rule" is known as "energy equivalence" (EER) because it predicts, in a fully saturated community, a similar use of resources and, therefore, productivity in different cohort sizes. This rule has recently received further support from metabolic ecology [see 7] but, at the same time, other analyses seemed to confute it [23].

We demonstrate that the EER holds true because the total resource use by a single cohort of the *same h* equals the number of trees of that $h$ class multiplied by its metabolic activity, i.e.: $E_{tot,h} = (-dP^>(h|h_c)/dh)*B_h$ and thus $E_{tot,h} = h^{-(1+2H)} * h^{(1+2H)} = 1$ when $h<h_c$, showing that the energy used is invariant with tree size only when tree height is considered as the independent variable and is lower than $h_c$. However we conclude that the energy equivalence holds *if and only if* the tree height is used as an independent variable and, further, that the range in which this rule might be true is limited by the range of the power law distribution. The reason the EER does not hold using the tree diameter, $D$, as the variable is that the correct variable transformation is $p(D)=p(h(D)) \ dh(D)/dD$, where $p(D)=-dP^>(D)/dD$, and $h(D)$ is the dependence of the height of a tree on its diameter, as pointed out by Stegen and White [27]. This transformation gives the scaling of $p(D) \propto D^{-7/3}$ in the power law regime. Because $B$ is believed to scale with $D^2$, at least for trees in tropical forests, the energy equivalence does not hold for grouping of individuals in diameter classes. In addition, given that the EER has its range of validity within the power law regime, it follows that comparing the total leaf area in very young (i.e. $h < h_{inf}$) or in relatively old stands (i.e. $h > h_c$) might lead to the result that the leaf area changes with size. This, however, might not disprove the EER principle as proposed by [23].

*Estimating the degree of disturbance*

As highlighted by Kerkhoff & Enquist [3], the scaling approach might be used as a tool for explaining the deviation from the expected of the tree-size distribution exponent in terms of ecosystem recovery. They considered the diameter as the independent variable and supported this idea by showing how the slopes of recently disturbed ecosystems were, in general, less steep than the others (almost undisturbed) (figure 3). Overall, our data are consistent with this idea. The new elements are: i) the self-thinning exponent can *change* depending on the community being considered and is not universally equal to -7/3 (if diameter is used as independent variable) or -3 (if tree height is used); ii) this slope must be calculated *only* in the range of power law behaviour excluding all trees beyond or near the cut-off. Indeed, relevant deviations from -2 have been reported, as in Coomes *et al.* [13], with estimated slopes more negative than -3 when the whole range of tree diameters was considered.

A comparison between the slope of the *potential* (i.e. steady state) self-thinning line and the *observed* tree-size distribution can provide a useful diagnostic tool to assess the legacy of large-scale and/or significant disturbances occurring within forest communities. Indeed, two of the sites (2 and 4) clearly showed the effects of human activity by the low number of trees, especially those of small diameter [28]. In both cases, the range of the power law behaviour of the self-thinning line is miniscule or, in other words, the exponent of the predicted tree-size distribution (i.e. $1+2H$) cannot be used to successfully fit the observed distributions: this suggests that the communities are non-saturated and that resources are not fully used.

In contrast, the $I_r$ of the two more natural forests (1 and 3) differed slightly but, in these cases, the range of fit of equation 2 is much larger.

More studies will be needed for a better understanding of the resilience of such communities, taking into account, for example, that in tropical forests the recovery time scale from a generic disturbance might be from several centuries until few thousand years [29].

Furthermore, attention must be paid when considering the minimal spatial dimension to reliably test the structural features of a community. In fact, if the disturbances (such as crown fires or wind-throws) affect large areas, plot sizes or their number must be selected accordingly in order to sample all size classes. In fact, the typical reverse J-shape distribution can appear when all size classes are thoroughly sampled, if necessary considering different even-aged stands with different sizes [13].

We also underline that our model differs from and is simpler than other models aiming to predict forest dynamics (e.g. the method proposed by Strigul *et al.* [30]). Indeed we can predict *only* the community structure (i.e. the relative variation of trees in the different size classes) in a virtual steady state condition, i.e. when resources are fully used.

Nonetheless, our approach can be useful in quantitatively assessing the degree of old-growthness in a forest. This has been tackled, until now, by using some structural forest attributes, for example the presence of dead wood, logs or snags [6]. Zenner [31] showed clearly that tree-size distribution curves evolve from the mature to old-growth condition: knowing the potential slope, it should be possible to quantify the transient recovery of the system.

**Conclusion**

Our approach shows that: i) it is possible to define a set of coherent relationships linking the structure and functionality at tree level based on a single parameter ($H$) and employ the power of finite size scaling. With a few plausible assumptions at community level, this approach might be used for successfully predicting the exponent of the self-thinning line [$P_h(h) \propto h^{-(1+2H)}$] thus suggesting that individual-based processes (in particular the scaling of crown volume) essentially drive the whole stand structure; ii) differently from previous theories, the scaling exponent ($H$) can change, allowing distinct slopes of the thinning lines, as frequently observed in different forest communities thus reconciling different results; iii) the predicted exponents, derived at tree level, represent the *potential* slope of the self-thinning line in the assumption of full resource use;

therefore any deviation observed in the distribution of trees can be used as a diagnostic tool for assessing ecosystem recovery since disturbance.

## Acknowledgements

This paper is warmly dedicated to Lucio Susmel, Emeritus Professor in Ecology at the University of Padova who was the first in Italy to promote the use of tree height for predicting forest structure and productivity. We thank Cariparo Foundation for financial support, the forest administration of the Provincia Autonoma di Bolzano - Alto Adige and the Regole d'Ampezzo for all the supports during the field activities. We also thank Silvia Lamedica and Alessandro Tenca for technical help in measuring stand structure. The research was also funded by the project UNIFORALL (University of Padova, Progetti di Ricerca di Ateneo CPDA110234).

## Appendix

The following example illustrates the role of finite "resources' by considering the random filling of a square of size $LxL$ with discs whose diameters are randomly extracted from, e.g., a uniform distribution in the interval (1,$D$) where $D$ is the maximum allowed diameter. A disc with random diameter is picked and added in the square in a randomly selected position if it does not overlap with the pre-existing discs. Otherwise it is discarded and the process is repeated until coverage of at least 90% is obtained. The figure 4 shows a log-log plot of the Cumulative Distribution Function for three cases corresponding to D=10, 100 and 300 with $L$=10$D$. As expected, there is a power law decay regime, with an exponent $\alpha$ – the power law regime is wider as $D$ increases. Note the sharp fall-off as the upper cut-off $D$ is approached. This suggests that the finite size scaling hypothesis for the conditional probability distribution of the diameters, $d$, given that the maximum allowed value is $D$, ought to have the following form

$$p_d\big(d|D\big) = d^{-\alpha} f_d\big(d/D\big) \qquad\qquad \text{(eqn 1b)}$$

or equivalently the corresponding CDF

$$P_d^{\,>}\big(d|D\big) = d^{-\alpha+1} F_d\big(d/D\big) \qquad\qquad \text{(eqn 2b)}$$

where $\alpha$ is the power law exponent and $f_d$ ($F_d$) is a function that takes into account the residual dependence on both $d$ and $D$ through their ratio rather than a generic separate dependence of $d$ and $D$. Note that $f_d$ and $F_d$ are related by the following equation:

$$f_d(d/D) = \big(\alpha - 1\big) F_d\big(d/D\big) - d/D\, F_d^{\,'}\big(d/D\big)$$ where $F'$ indicates the first derivative of $F$. If $d << D$ one expects that the (cumulative) probability distribution is not affected by the cut-off and so it is plausible that $f_d(d/D)\big(F_d(d/D)\big)$ tends to a constant in this regime. On the other hand since there are no discs with diameter larger than $D$, we expect that $f_d(d/D)\big(F_d(d/D)\big)$ decays rapidly when $d$ approaches $D$. This hypothesis, i.e. that the $d$ and $D$ dependence of the cumulative probability distribution takes the form equation 2b is easily verified by re-plotting the data for $d^{\alpha-1} P_d^{\,>}\big(d|D\big)$ versus $d/D$. According to equation 2b this should give a unique function $F_d(d/D)$ rather than different functions for different values of $D$. The quality of the *collapse* in the inset clearly shows that the hypothesis equation 2b, as well as equation 1b, are well-grounded.

| SITE name | Trees ha$^{-1}$ | Max tree height (m) | Dominant tree height (m) | Mean DBH (cm) | Max DBH (cm) | Stand BA (m$^2$ ha$^{-1}$) | Number of trees |
|---|---|---|---|---|---|---|---|
| 1 - Slatioara | 1281 | 50.0 | 44.0 | 24.1 | 143 | 58.2 | 5124 |
| 2 - Cansiglio | 623 | 42.1 | 37.1 | 30.8 | 95 | 46.5 | 2492 |
| 3 - Cortina d'Ampezzo | 791 | 26.0 | 23.5 | 20.0 | 95 | 24.8 | 3164 |
| 4 - Obereggen | 489 | 35.3 | 30.0 | 33.2 | 75 | 42.2 | 1956 |

Table 1

Table 2

| SITE | $h_c$ | $h_{inf}$ | $I_r$ | $R^2$ fitting | N. trees |
|---|---|---|---|---|---|
| 1 - Slatioara | 33.16±0.11 | 5.99±0.01 | 0.62 | 0.999 | 2991 |
| 2 - Cansiglio | 30.14±0.06 | 16.76±0.06 | 0.36 | 0.998 | 834 |
| 3 - Cortina d'Ampezzo | 20.30±0.26 | 3.74±0.03 | 0.70 | 0.988 | 1508 |
| 4 - Obereggen | 24.01±0.03 | 24.50±0.12 | -0.02 | 0.999 | 0 |

**Table and figure legends**

Table 1.  Main structural characteristics of the four sampled forests (BA= basal area)

Table 2. Results of the least-squares fits of the cumulative distributions of tree heights in the four sites using equation 2. The values of the "recovery index" ($I_r$) indicate the relative amplitude of the range of the power law behaviour ($h_c$-$h_{inf}$) in the CDFs (± 95% confidence intervals). Values of $I_r <$ 0.5 would suggest a more recent disturbance. The last column, N, denotes the number of trees within the interval $h_c$-$h_{inf}$ and therefore differs from the value in table 1.

Figure 1. Scaling of crown volume (calculated as $r_{cro}^2 * l_{cro}$) in the four measured sites (above sites 1,2; below sites 3,4) . The ±95% confidence intervals of the $H$ estimations are shown in parenthesis.

Figure 2.  Scaling of crown volume (calculated as $r_{cro}^2 * l_{cro}$) in different species of the same site. Site 1 (top) *Abies alba* (AA) and *Fagus sylvatica*  (FS); site 3 (bottom) *Larix decidua*, (LD) and *Pinus cembra* (PC).

Figure 3. Cumulative distribution function (CDF) of the tree heights in the four sites and fitting curves used to estimate $h_{inf}$ and $h_c$ [eqn (2)] in normal scale (above) and in a log-log plot (below).

Figure 4. Appendix.  Numerical example of the finite size scaling.

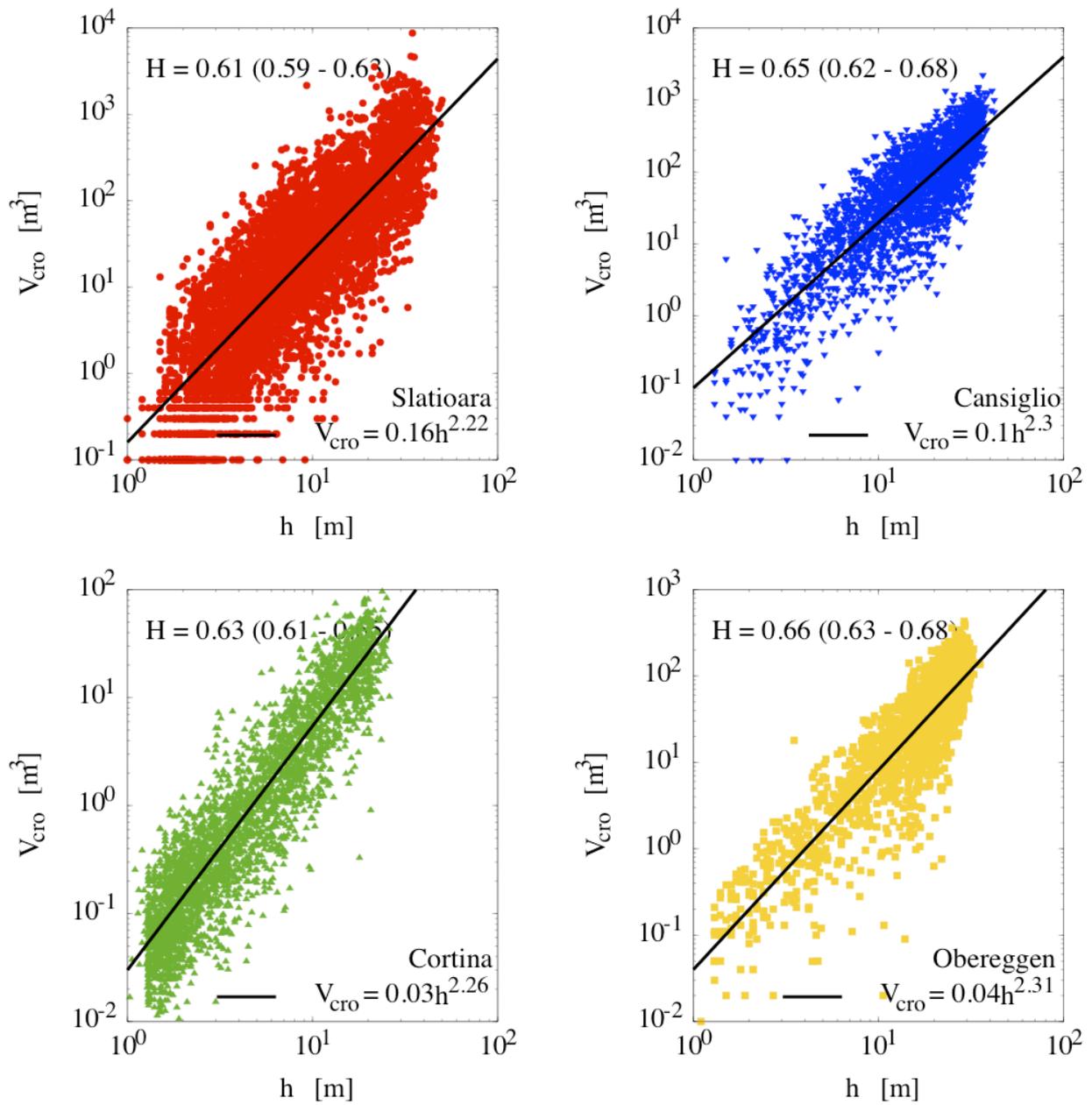

**Figure 1**

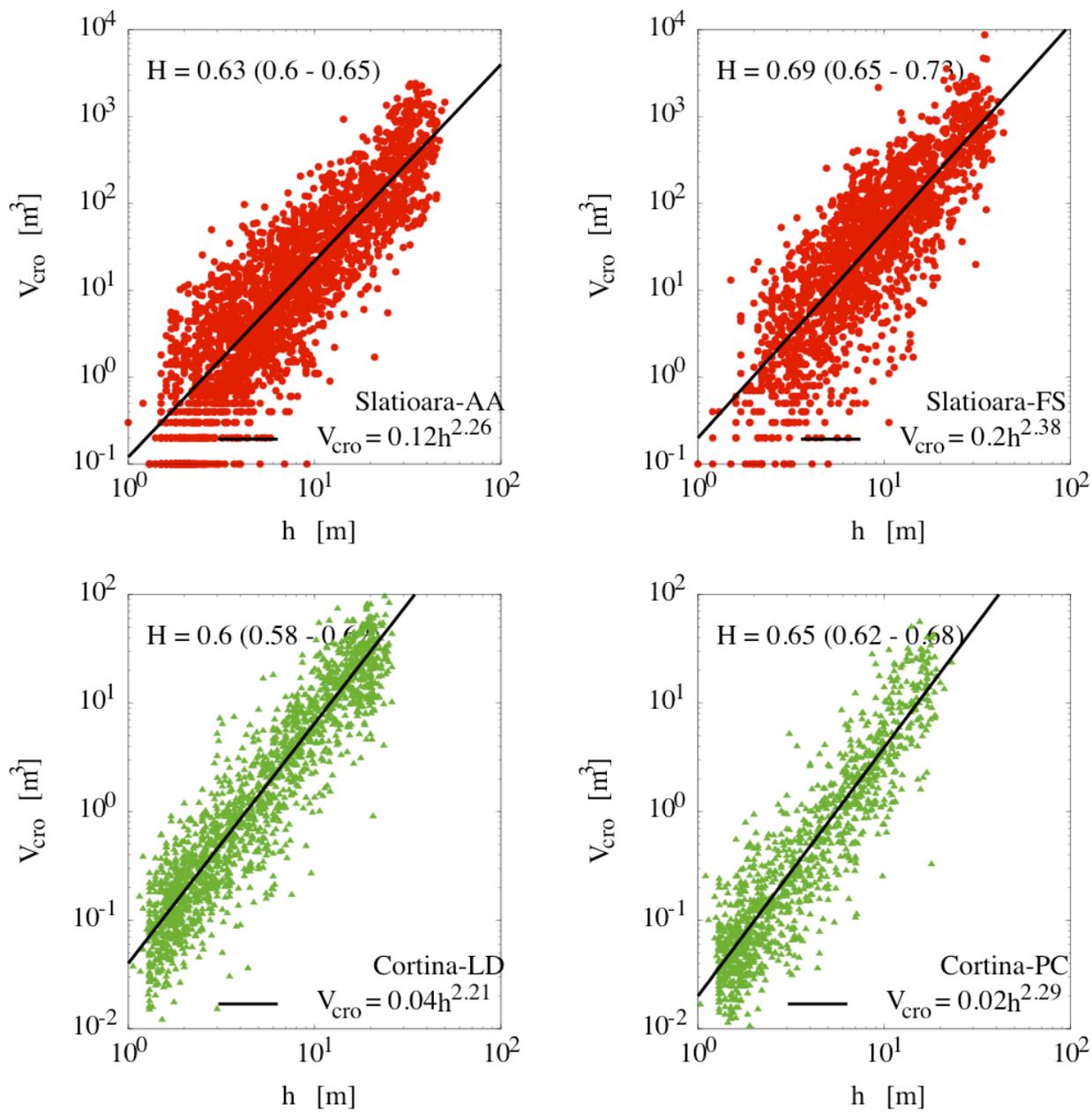

Figure 2

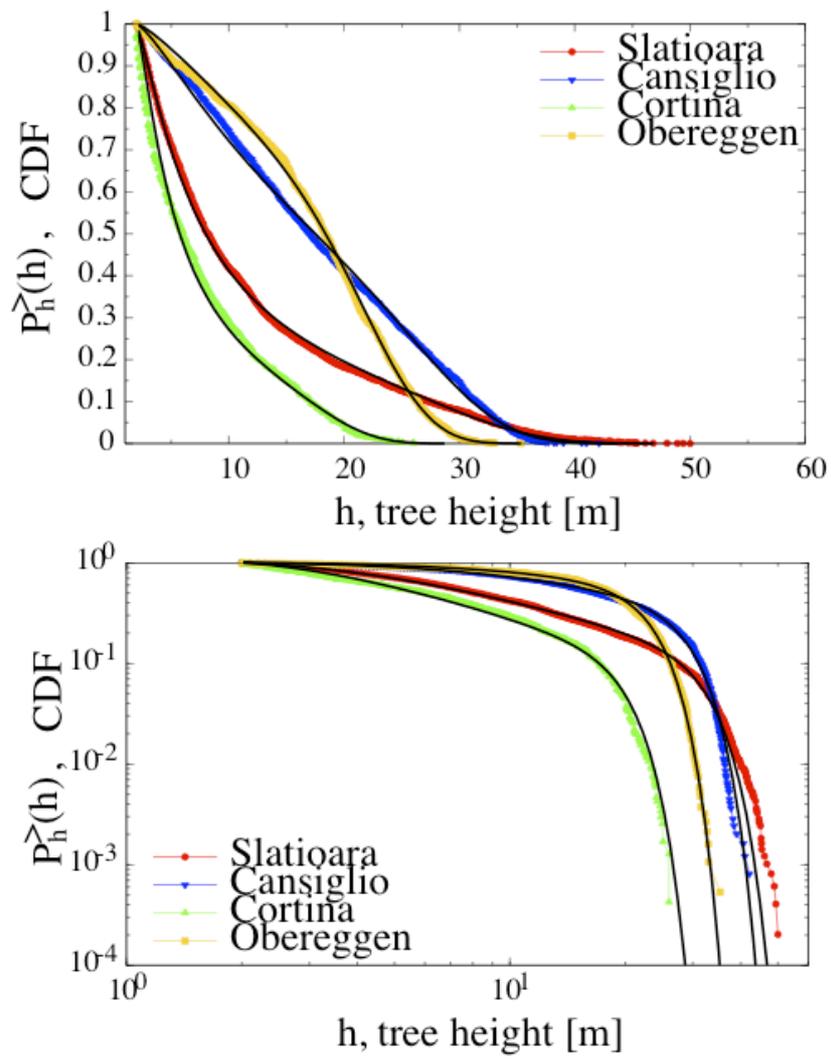

**Figure 3**

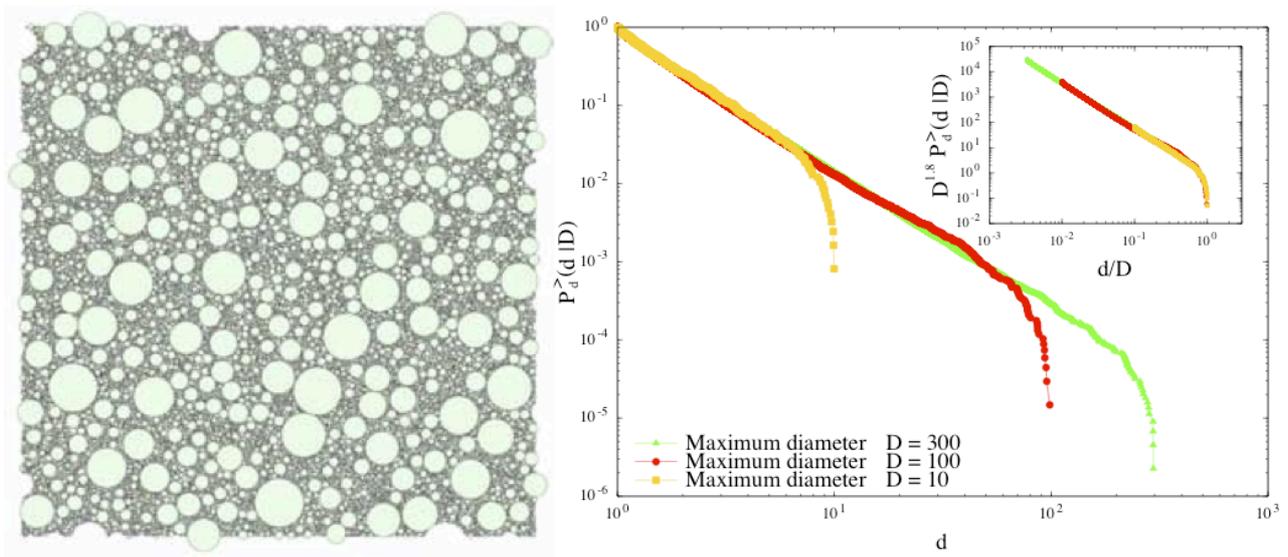

**Figure 4**